\documentstyle[preprint,eqsecnum,aps]{revtex}
\begin{document}
\title{One-Loop Divergences in Simple Supergravity: 
Boundary Effects}
\author{Giampiero Esposito$^{1,2}$\thanks
{Electronic address: esposito@napoli.infn.it}
and Alexander Yu. Kamenshchik$^{3}$\thanks
{Electronic address: grg@ibrae.msk.su}}
\address{${ }^{1}$ Istituto Nazionale di Fisica Nucleare,
Sezione di Napoli,\\
Mostra d'Oltremare Padiglione 20, 80125 Napoli, Italy\\
${ }^{2}$ Dipartimento di Scienze Fisiche,\\
Mostra d'Oltremare Padiglione 19, 80125 Napoli, Italy\\
${ }^{3}$ Nuclear Safety Institute, Russian Academy of Sciences,\\
52 Bolshaya Tulskaya, Moscow 113191, Russia}
\maketitle
\begin{abstract}
This paper studies the semiclassical approximation of simple
supergravity in Riemannian four-manifolds with boundary, within
the framework of $\zeta$-function regularization.
The massless nature of gravitinos, jointly with the presence of a
boundary and a local description in terms of potentials for spin
${3\over 2}$, force the background to be totally flat. First,
nonlocal boundary conditions of the spectral type are imposed on
spin-${3\over 2}$ potentials, jointly with boundary conditions
on metric perturbations which are completely invariant under
infinitesimal diffeomorphisms. The axial gauge-averaging functional
is used, which is then sufficient to ensure
self-adjointness. One thus finds that the contributions of ghost and
gauge modes vanish separately. Hence the contributions to the one-loop
wave function of the universe reduce to those $\zeta(0)$ values
resulting from physical modes only. Another set of mixed boundary
conditions, motivated instead by local supersymmetry and first proposed 
by Luckock, Moss and Poletti, is also analyzed. In this case the
contributions of gauge and ghost modes do not cancel each other.
Both sets of boundary conditions lead to a nonvanishing $\zeta(0)$
value, and spectral boundary conditions are also studied when two
concentric three-sphere boundaries occur.
These results seem to point out that simple 
supergravity is not even one-loop finite in the
presence of boundaries. 
\end{abstract}

\section{Introduction}

Although Einstein's general relativity leads to a theory of
quantum gravity which is not perturbatively renormalizable
[1], the analysis of the semiclassical approximation remains
of crucial importance to test the internal consistency of any
theory of quantum gravity. For this purpose, it is necessary to
achieve a thorough understanding of the problem of boundary
conditions in the theory of quantized fields. 
Indeed, the path-integral representation of
the propagator [2], the general theory of the effective action
[3], and the recent attempts to define a quantum state of the
universe [4--6], provide three relevant examples where the
appropriate formulation of boundary conditions plays a crucial
role to obtain a well defined model of some properties of 
quantum gravity.

For gauge fields and gravitation, one may reduce the theory to 
its physical degrees of freedom by imposing a gauge condition
{\it before} quantization [7,8], or one may use the Faddeev-Popov
formalism for quantum amplitudes [9], 
or the extended-phase-space
Hamiltonian formalism of Batalin, Fradkin and Vilkovisky [10].
Moreover, a powerful nondiagrammatic method to perform the
one-loop analysis is the one which relies on $\zeta$-function
regularization [11]. This is a naturally occurring technique, since
semiclassical amplitudes involve definition and calculation of
determinants of elliptic, self-adjoint differential operators.

Once these choices are made, there are still many problems which
deserve a careful consideration. They are as follows.

(i) {\it Choice of background four-geometry}. This may be flat
Euclidean four-space, which is relevant for massless theories
[12], or the de Sitter four-sphere [13], which is relevant for
inflationary cosmology [14], or more general curved backgrounds.
The latter appear interesting for a better understanding of 
quantum field theory in curved space-time.

(ii) {\it Choice of boundary three-geometry}. This may consist
of two three-surfaces (e.g. two concentric three-spheres),
motivated by quantum field theory [15], or just one three-surface
(e.g. one three-sphere), motivated by quantum cosmology [4,5], 
or more complicated examples of boundary three-geometries.

(iii) {\it Choice of gauge-averaging functional}. For example,
one may study Lorentz or Coulomb gauge for Euclidean Maxwell
theory [16--18], or de Donder or axial gauge for gravitation
[19--22], or the noncovariant gauges 
proposed in Refs. [16,18,23], which
take explicitly into account extrinsic-curvature effects.

(iv) {\it Boundary conditions}. These may be Dirichlet, Neumann
or Robin for scalar fields [24]; local or spectral for massless
spin-${1\over 2}$ fields [25-31]; magnetic or electric for
Maxwell theory [16--18,23]; local or spectral for spin-${3\over 2}$
potentials [26--31]; completely invariant under infinitesimal 
diffeomorphisms on metric perturbations [32], or expressed in
terms of (complementary) projectors at the boundary [33], or
Robin on spatial perturbations and Dirichlet on normal
perturbations [21], or manifestly symmetric [34], for linearized
gravity.

Since the detailed calculations in Refs. 
[16--31] have led to the
correct values of the one-loop divergences for spin-${1\over 2}$
fields, Euclidean Maxwell theory and Euclidean quantum gravity for
various choices of boundary conditions, including all ghost and
gauge modes whenever appropriate, 
the last open problem in this respect was the
evaluation of one-loop divergences in simple supergravity in the
presence of boundaries. Our paper is thus devoted to a detailed
investigation of such an issue.

Simple supergravity is the simplest supersymmetric gauge theory
of gravitation [35]. Its action functional consists of the
Einstein-Hilbert action, the massless gravitino action [36]
\begin{equation}
I_{\psi,{\widetilde \psi}} \equiv {1\over 2}
\int_{M} \epsilon^{\mu \nu \rho \sigma} 
\left[{\widetilde \psi}_{\; \; \; \mu}^{A'} \;
e_{AA' \nu} \; D_{\rho} \; \psi_{\; \; \sigma}^{A}
-\psi_{\; \; \mu}^{A} \; e_{AA' \nu} \; D_{\rho} \;
{\widetilde \psi}_{\; \; \; \sigma}^{A'} \right] d^{4}x \; ,
\end{equation}
jointly with real and complex auxiliary fields (which
are necessary to close the supersymmetry algebra) and suitable 
boundary terms depending on the choices of boundary conditions 
at the bounding three-surfaces. With our notation, which 
relies on Refs. [36,37], the gravitino potential is represented
by the pair of anticommuting, independent (i.e., not related
by any conjugation) spinor-valued one-forms 
$\psi_{\; \; \mu}^{A}$ and ${\widetilde \psi}_{\; \; \; \mu}^{A'}$.
These are obtained by contraction of spinor fields with the spinor
version of the tetrad as [12]
\begin{equation}
\psi_{A \; \mu}=\Gamma_{\; \; \; AB}^{C'} \;
e_{\; \; C' \; \mu}^{B} \; ,
\end{equation}
\begin{equation}
{\widetilde \psi}_{A' \; \mu}=\gamma_{\; \; A'B'}^{C} \;
e_{C \; \; \; \mu}^{\; \; B'} \; .
\end{equation}
Further details about spinor formalism can be found in our 
Appendix A, which relies again on Refs. [12,36,37].

The spinor fields $\Gamma$ and $\gamma$ occurring in Eqs. (1.2) and
(1.3) are referred to as Rarita-Schwinger 
potentials [12,38,39]. They are
subject to the {\it infinitesimal} gauge transformations [12]
\begin{equation}
{\widehat \Gamma}_{\; \; \; BC}^{A'} \equiv 
\Gamma_{\; \; \; BC}^{A'}+\nabla_{\; \; \; B}^{A'} \; \nu_{C} \; ,
\end{equation}
\begin{equation}
{\widehat \gamma}_{\; \; B'C'}^{A} \equiv 
\gamma_{\; \; B'C'}^{A}+\nabla_{\; \; B'}^{A} \; \mu_{C'} \; .
\end{equation}
For the spinor fields $\nu_{C}$ and $\mu_{C'}$ to be freely
specifiable inside the background four-manifold, the trace-free
part of the Ricci tensor has to vanish, jointly with the scalar
curvature. Hence the background is forced to be 
Ricci-flat [12,40--42].

Further restrictions are obtained on considering a {\it local}
description of $\Gamma$ and $\gamma$ in terms of a second 
potential [12,38,39]. For example, on expressing locally the 
Rarita-Schwinger potentials as [12,38,39]
\begin{equation}
\gamma_{A'B'}^{\; \; \; \; \; \; C}=\nabla_{BB'} \;
\rho_{A'}^{\; \; \; CB} \; ,
\end{equation}
\begin{equation}
\Gamma_{AB}^{\; \; \; \; \; C'} = \nabla_{BB'} \; 
\theta_{A}^{\; \; C'B'} \; ,
\end{equation}
one finds that the basic equations obeyed by the second potentials
$\rho$ and $\theta$ are invariant under infinitesimal gauge
transformations with gauge fields $\omega^{D}$ and $\sigma^{D'}$
if and only if [12]
\begin{equation}
\psi_{AFLD} \; \omega^{D}=0 \; ,
\end{equation}
\begin{equation}
{\widetilde \psi}_{A'F'L'D'} \; \sigma^{D'}=0 \; .
\end{equation}
With a standard notation, $\psi_{AFLD}$ and 
${\widetilde \psi}_{A'F'L'D'}$ are the anti-self-dual and
self-dual parts of the Weyl curvature spinor, respectively. 
Thus, to ensure unrestricted gauge freedom (except at the boundary)
for the second potentials $\rho$ and $\theta$, one is forced to
work in totally flat Euclidean backgrounds.
Such a restriction for gravitinos results from the form of the
action (1.1), which involves both $\psi_{\; \; \mu}^{A}$ and
${\widetilde \psi}_{\; \; \; \mu}^{A'}$, jointly with the local
description (1.6) and (1.7) and the form of any admissible set of 
boundary conditions.

Of course, the operator acting on Rarita-Schwinger
potentials is a Dirac operator. Within our framework,
which deals with positive-definite four-metrics, the Dirac operator
is a first-order elliptic operator which maps primed spinor fields
to unprimed spinor fields, and the other way around [39,43]. Thus,
the specification of the whole gravitino potential at the boundary
would lead to an over-determined problem. One thus has a choice of
spectral or local boundary conditions for Rarita-Schwinger
potentials, and they are both studied in our paper in the presence
of three-sphere boundaries [25,26].

Spectral conditions reflect a choice which leads to a well posed
classical boundary-value problem. In other words, the massless
Rarita-Schwinger potential subject to gauge conditions and linearized
supersymmetry constraints is splitted into a regular part and a
singular part. The regular part is an infinite sum of modes 
multiplying harmonics having positive eigenvalues of the intrinsic
three-dimensional Dirac operator of 
the boundary [26,31]. By contrast,
the singular part is an infinite sum of modes multiplying harmonics
having negative eigenvalues of the intrinsic three-dimensional 
Dirac operator of the boundary [26,31]. Such an identification relies
therefore on a {\it nonlocal} operation, i.e., the separation of the
spectrum of a first-order elliptic operator into its positive and
negative part [43,44].

When the corresponding semiclassical approximation of quantum theory
is studied, only half of the gravitino potential is set to
zero at the boundary. Bearing in mind the scheme described above,
and with the notation of Ref. [31], one writes the spectral boundary
conditions on gravitino perturbations in the form
\begin{equation}
\Bigr[\psi_{i(+)}^{A}\Bigr]_{\partial M}=0 \; ,
\end{equation}
\begin{equation}
\Bigr[{\widetilde \psi}_{i(+)}^{A'}\Bigr]_{\partial M}=0 \; ,
\end{equation}
where the label $(+)$ denotes the part of the perturbation potential
corresponding to the regular part of the underlying classical 
theory. The issue of boundary conditions on $\psi_{0}^{A}$ and
${\widetilde \psi}_{0}^{A'}$ will be discussed in Sec. III.
To ensure invariance of the boundary conditions (1.10) and (1.11)
under the infinitesimal gauge transformations (1.4) and (1.5),
we require that
\begin{equation}
\Bigr[\nabla_{\; \; \; B}^{A'} \; {\nu_{C}}_{(+)}\Bigr]_{\partial M}
=0 \; ,
\end{equation}
\begin{equation}
\Bigr[\nabla_{\; \; B'}^{A} \; {\mu_{C'}}_{(+)}\Bigr]_{\partial M}
=0 \; .
\end{equation}

As far as metric perturbations $h_{\mu \nu}$ are concerned, we
are interested in setting to zero at the boundary the spatial
perturbations [32], i.e., 
\begin{equation}
\Bigr[h_{ij}\Bigr]_{\partial M}=0 \; .
\end{equation}
The six boundary conditions (1.14) are invariant under the
infinitesimal gauge transformations [19,20,22]
\begin{equation}
{ }^{(\varphi)}h_{\mu \nu} \equiv h_{\mu \nu}
+\nabla_{(\mu} \; \varphi_{\nu)} \; ,
\end{equation}
if the whole ghost one-form $\varphi_{\nu}$ is set to zero
at the boundary:
\begin{equation}
\Bigr[\varphi_{\nu}\Bigr]_{\partial M}=0 \; .
\end{equation}
At this stage, the boundary conditions on the normal components
$h_{00}$ and $h_{0i}$ are invariant under (1.15) if and only
if the whole gauge-averaging functional $\Phi_{\mu}$ is set to
zero at the boundary:
\begin{equation}
\Bigr[\Phi_{\mu}(h)\Bigr]_{\partial M}=0 \; .
\end{equation}
As shown in Ref. [22], a choice of gauge-averaging
functional which leads to self-adjoint operators on metric
perturbations is then the {\it axial}, i.e.,
\begin{equation}
\Phi_{\mu}(h) \equiv n^{\rho} \; h_{\mu \rho} \; ,
\end{equation}
where $n^{\rho}$ is the normal to the boundary. Since also the
proof of essential self-adjointness of squared Dirac operators
with spectral boundary conditions has been recently put on
solid ground [45], we have described so far a scheme where the boundary
conditions are completely invariant under infinitesimal gauge
transformations, the boundary operators on metric perturbations are
local and the differential operators on perturbative modes are
self-adjoint. Of course, the boundary conditions (1.14) and (1.17)
can be re-expressed in terms of tetrad vectors, and the one-loop
results coincide with those obtained from the metric 
formulation.

Another set of boundary conditions is instead motivated by the
work in Ref. [46], where it is shown that the spatial tetrad 
$e_{\; \; \; \; \; i}^{AA'}$ and the projection
$\Bigr(\pm {\widetilde \psi}_{i}^{A'}-\sqrt{2} \;
{_{e}n_{A}^{\; \; A'}} \; \psi_{i}^{A}\Bigr)$ transform into
each other under half of the local supersymmetry transformations 
at the boundary. The resulting boundary conditions in one-loop
quantum cosmology about flat Euclidean four-space bounded by a
three-sphere with Euclidean normal ${_{e}n_{A}^{\; \; A'}}$
(see (A10)) take the form [26]
\begin{equation}
\sqrt{2} \; {_{e}n_{A}^{\; \; A'}} \; \psi_{i}^{A}
= \pm {\widetilde \psi}_{i}^{A'} \; {\mbox {at}} \;
\partial M \; ,
\end{equation}
jointly with the six boundary conditions (1.14) and the following
four boundary conditions on normal components of metric 
perturbations [19,20]:
\begin{equation}
\left[{\partial h_{00} \over \partial \tau}
+{6\over \tau}h_{00}-{\partial \over \partial \tau}
\Bigr(g^{ij}h_{ij}\Bigr)\right]_{\partial M}=0 \; ,
\end{equation}
\begin{equation}
\Bigr[h_{0i}\Bigr]_{\partial M}=0 \; ,
\end{equation}
where $g$ is the flat background four-metric, and $\tau$ is the
Euclidean-time variable, which plays the role of a radial coordinate.
Moreover, the ghost one-form for gravitons 
is subject to mixed boundary conditions,
in that the normal component $\varphi_{0}$ obeys Dirichlet
conditions [20]:
\begin{equation}
\Bigr[\varphi_{0}\Bigr]_{\partial M}=0 \; ,
\end{equation}
and the tangential components obey Robin conditions [20]:
\begin{equation}
\biggr[{\partial \varphi_{i}\over \partial \tau}
-{2\over \tau}\varphi_{i}\biggr]_{\partial M}=0 \; .
\end{equation}

Section II studies the ghost sector in the axial gauge for simple
supergravity, when spectral boundary conditions are imposed on
gravitino perturbations.
Section III is devoted to the corresponding graviton
and gravitino contributions. Section IV evaluates the one-loop
divergence with the Luckock-Moss-Poletti (LMP) boundary conditions
(1.14) and (1.19)--(1.23). Section V performs the one-loop
analysis in the axial gauge when two concentric three-sphere
boundaries occur, with the same boundary conditions of Secs.
II and III. Concluding remarks are presented in 
Sec. VI, and relevant details are given in the Appendices.

\section{Axial gauge: ghost sector} 

In this section we study the ghost sector of the gauge-fixed path
integral for simple supergravity in the axial gauge. We make such a
choice because, in the Barvinsky framework 
[32] for boundary conditions 
outlined in the first part of Sec. I, this is the only gauge
condition which ensures self-adjointness of the differential
operators acting on perturbative modes [22]. We begin with the
gravitino analysis, since the gravitational sector in the axial
gauge is thoroughly examined in Ref. [22].

The spinor fields $\Gamma_{\; \; \; AB}^{C'}$ and 
$\gamma_{\; \; A'B'}^{C}$ occurring in Eqs. (1.2) and (1.3) admit
the general decomposition into symmetric and antisymmetric parts
\begin{equation}
\Gamma^{ACC'}=\Gamma^{(AC)C'}+{1\over 2}\epsilon^{AC}
\; \Gamma_{D}^{\; \; DC'} \; ,
\end{equation}
\begin{equation}
\gamma^{A'C'C}=\gamma^{(A'C')C}+{1\over 2}\epsilon^{A'C'}
\; \gamma_{D'}^{\; \; \; D'C} \; .
\end{equation}
In the axial gauge (cf. Eqs. (3.13) and (3.14)), 
the following spinor fields are set to zero:
\begin{equation}
\Phi^{A}(\Gamma) \equiv \Gamma^{ACC'} \; {_{e}n_{CC'}}=0 \; ,
\end{equation}
\begin{equation}
{\widetilde \Phi}^{A'}(\gamma) \equiv 
\gamma^{A'C'C} \; {_{e}n_{CC'}}=0 \; .
\end{equation}
This is the purely two-spinor formulation of the gauge conditions
studied in Refs. [26,31]. Since the symmetric part of $\Gamma$
may be further split as the sum of two spinor fields having 
vanishing contraction with ${_{e}n_{CC'}}$ [31], and a third
contribution proportional to [37]
$$
\Gamma_{D \; \; F'}^{\; \; D} \; {_{e}n^{F'(A}} \;
{_{e}n^{C)C'}} \; ,
$$
and similarly for the $\gamma$-potential, the gauge conditions
(2.3) and (2.4) set to zero the traces $\Gamma_{D}^{\; \; DF'}$
and $\gamma_{D'}^{\; \; D'F}$ of the Rarita-Schwinger potentials.

In the quantum theory via path integrals, however, one performs
a {\it gauge-averaging} [16] involving the left-hand 
sides of (2.3) and (2.4).
The resulting ghost operators are obtained by studying the
transformation of $\Phi^{A}$ and ${\widetilde \Phi}^{A'}$ under
the infinitesimal gauge transformations (1.4) and (1.5).
Hence one finds the ghost operator
\begin{equation}
{\cal D}_{C}^{\; \; A} \equiv {_{e}n_{CC'}} \nabla^{AC'} \; ,
\end{equation}
acting on the spinor field $\nu^{C}$, and the ghost operator
\begin{equation}
{\cal F}_{C'}^{\; \; \; A'} \equiv {_{e}n_{CC'}} \nabla^{CA'} \; ,
\end{equation}
acting on the spinor field $\mu^{C'}$. These are four-dimensional
operators but, unlike the four-dimensional Dirac operator, they
map elements of a spin-space (either primed or unprimed) to
elements of the same spin-space.

The ghost fields are now expanded in harmonics on a family of
three-spheres centred on the origin as
\begin{equation}
\nu^{A}=\sum_{n=0}^{\infty}\sum_{p,q=1}^{(n+1)(n+2)}
\alpha_{n}^{pq}\Bigr[m_{np}(\tau)\rho^{nqA}
+{\widetilde r}_{np}(\tau){\overline \sigma}^{nqA}\Bigr] \; ,
\end{equation}
\begin{equation}
\mu^{A'}=\sum_{n=0}^{\infty}\sum_{p,q=1}^{(n+1)(n+2)}
\alpha_{n}^{pq}\Bigr[{\widetilde m}_{np}(\tau)
{\overline \rho}^{nqA'}+r_{np}(\tau)\sigma^{nqA'}\Bigr] \; ,
\end{equation}
where the $\alpha_{n}$ are block-diagonal matrices with blocks
$\pmatrix {1&1 \cr 1&-1 \cr}$, and the harmonics, defined in
Ref. [47], obey the eigenvalue equations described in Appendix A.
However, unlike Ref. [47], different
conventions for numerical factors and the dependence on the
radial coordinate have been used, for convenience.
  
Given the eigenfunction expansions
\begin{equation}
\nu^{A}=\sum_{\lambda}{\cal C}_{\lambda}^{1} 
\; \nu_{(\lambda)}^{A} \; ,
\end{equation}
\begin{equation}
\mu^{A'}=\sum_{\tilde \lambda}{\cal C}_{\tilde \lambda}^{2} \;
\mu_{(\tilde \lambda)}^{A'} \; ,
\end{equation}
we now study the eigenvalue equations
\begin{equation}
{\cal D}_{C}^{\; \; A} \; \nu_{(\lambda)}^{C}
=\lambda \; \nu_{(\lambda)}^{A} \; ,
\end{equation}
\begin{equation}
{\cal F}_{C'}^{\; \; A'} \; \mu_{(\tilde \lambda)}^{C'}
={\tilde \lambda} \; \mu_{(\tilde \lambda)}^{A'} \; .
\end{equation}
These equations are supplemented by the boundary conditions
\begin{equation}
\Bigr[\nabla_{\; \; \; B}^{A'} \; {\nu_{C}}_{(\lambda)}(+)
\Bigr]_{\partial M}=0 \; ,
\end{equation}
\begin{equation}
\Bigr[\nabla_{\; \; B'}^{A} \; {\mu_{C'}}_{(\tilde \lambda)}(+)
\Bigr]_{\partial M}=0 \; ,
\end{equation}
which ensure the invariance of the spectral boundary conditions 
(1.10) and (1.11) under the infinitesimal gauge transformations
(1.4) and (1.5).

In our flat Euclidean background bounded by a three-sphere, the
left-hand side of Eq. (2.11) reduces to
\begin{equation}
{\cal D}_{C}^{\; \; A} \; \nu_{(\lambda)}^{C}
=-{1\over 2}{\partial \over \partial \tau}\nu_{(\lambda)}^{A}
-{_{e}n_{C'}^{\; \; \; (C}} \;
e^{A)C'i} \; { }^{(4)}\nabla_{i} {\nu_{C}}_{(\lambda)} \; ,
\end{equation}
by virtue of Eq. (A12). On using the Eqs. (A11), (A13) and (A14)
of Appendix A, this yields
\begin{equation}
{\cal D}_{C}^{\; \; A} \; \nu_{(\lambda)}^{C}
=-{1\over 2}\biggr({\partial \over \partial \tau}
\nu_{(\lambda)}^{A}- {3\over 2\tau}\nu_{(\lambda)}^{A}
\biggr) - {_{e}n_{C'}^{\; \; \; (C}} \;
e^{A)C'i} \; { }^{(3)}\nabla_{i} {\nu_{C}}_{(\lambda)} \; .
\end{equation}
Thus, apart from an unessential proportionality factor of
${1\over 2}$ which may be absorbed into the definition of the
operators (2.5) and (2.6), one finds the following eigenvalue
equations for the ghost modes occurring in the expansion (2.7)
(see (A15)--(A18)):
\begin{equation}
-\Bigr({d\over d\tau}+{n\over \tau}\Bigr)m_{np}
=\lambda_{n} \; m_{np} \; ,
\end{equation}
\begin{equation}
-\Bigr({d\over d\tau}-{(n+3)\over \tau}\Bigr)
{\widetilde r}_{np}=\lambda_{n} \; {\widetilde r}_{np} \; .
\end{equation}
Moreover, an entirely analogous treatment of Eq. (2.12)
yields the eigenvalue equations
for the second set of ghost modes in the form
\begin{equation}
-\Bigr({d\over d\tau}-{n\over \tau}\Bigr)
{\widetilde m}_{np}={\widetilde \lambda}_{n} \;
{\widetilde m}_{np} \; ,
\end{equation}
\begin{equation}
-\Bigr({d\over d\tau}+{(n+3)\over \tau}\Bigr)r_{np}
={\widetilde \lambda}_{n} \; r_{np} \; .
\end{equation}
The solutions of Eqs. (2.17)--(2.20) can be written as
\begin{equation}
m_{np}(\tau,\lambda_{n})=\tau^{-n} \; e^{-\lambda_{n}\tau}
\; m_{np}^{0} \; ,
\end{equation}
\begin{equation}
{\widetilde r}_{np}(\tau,\lambda_{n})=\tau^{n+3}
\; e^{-\lambda_{n}\tau} \; {\widetilde r}_{np}^{0} \; ,
\end{equation}
\begin{equation}
{\widetilde m}_{np}(\tau,{\widetilde \lambda}_{n})
=\tau^{n} \; e^{-{\widetilde \lambda}_{n} \tau}
\; {\widetilde m}_{np}^{0} \; ,
\end{equation}
\begin{equation}
r_{np}(\tau,{\widetilde \lambda}_{n})
=\tau^{-(n+3)} \; e^{-{\widetilde \lambda}_{n} \tau}
\; r_{np}^{0} \; .
\end{equation}
Note that, unlike the case of a massless spin-${1\over 2}$
field, regularity at the origin makes it necessary to set to
zero {\it everywhere} the modes $m_{np}$ and $r_{np}$.

Contraction of the boundary conditions (2.13) and (2.14) with
the Euclidean normal, jointly with the identity in flat
Euclidean four-space
\begin{equation}
{_{e}n_{BA'}}\nabla^{BA'} \; \nu_{(\lambda)}^{C}
=-{\partial \over \partial \tau} \nu_{(\lambda)}^{C} \; ,
\end{equation}
and with the Eqs. (2.22) and (2.23) for the regular modes, is
thus found to lead to the discrete spectra
\begin{equation}
\lambda_{n}={(n+3)\over a} \; \; \; \; \forall n \geq 0 \; ,
\end{equation}
\begin{equation}
{\widetilde \lambda}_{n}={n\over a} \; \; \; \; \forall n \geq 0 \; ,
\end{equation}
where $a$ is the three-sphere radius, hereafter set to 1 for
convenience, since it does not affect the $\zeta(0)$ value.
This implies that a first-order, elliptic and
positive-definite operator $\cal A$ exists with spectrum (2.26),
for which a $\zeta$-function can be defined as
\begin{equation}
\zeta_{\cal A}(s) \equiv \sum_{n=0}^{\infty}
(n+1)(n+2)(n+3)^{-s} \; ,
\end{equation}
and a first-order, elliptic and nonnegative operator $\cal B$
exists with spectrum (2.27) and {\it finite-dimensional} 
null-space, for which a $\zeta$-function can be defined as
\begin{equation}
\zeta_{\cal B}(s) \equiv 2 + \sum_{n=1}^{\infty}
(n+1)(n+2)n^{-s} \; ,
\end{equation}
where the dimension of the null-space has been included in the
definition of the $\zeta$-function, following Ref. [48]
(for the opposite view in the literature, see e.g. Ref. [49]).
Note that our definitions (2.28) and (2.29) are particular cases of 
$\eta$-functions [31,43,48], since our operators are first-order but
either positive-definite or at least nonnegative.
By using Eqs. (B3)--(B6) one thus finds
\begin{equation}
\zeta_{\cal A}(0)=\zeta_{H}(-2,3)-3\zeta_{H}(-1,3)
+2 \zeta_{H}(0,3)=-{3\over 4} \; ,
\end{equation}
\begin{equation}
\zeta_{\cal B}(0)=2+\zeta_{R}(-2)+3\zeta_{R}(-1)
+2 \zeta_{R}(0)={3\over 4} \; .
\end{equation}
Hence the ghost gravitino contribution to the one-loop
divergence in the axial gauge vanishes in our flat background
bounded by a three-sphere:
\begin{equation}
\zeta_{\mbox {ghost}}(0)=\zeta_{\cal A}(0)+\zeta_{\cal B}(0)=0 \; .
\end{equation}
Note that no further ghost fields occur in our calculation
(cf. Sec. IV and Appendix A), since the axial gauge already
has the effect to reduce the linearized gravitino potential
to its two physical degrees of freedom, corresponding to
helicities $3/2$ and $-3/2$ (see also Sec. III).

The ghost analysis for gravitons in the axial gauge (1.18) is
performed in detail in Ref. [22], and we only need to say that
the ghost operator turns out to be [22]
\begin{equation}
{\cal F}_{\mu}^{\; \; \nu}=\delta_{\mu}^{\; \; \nu}
\; n^{\rho}\nabla_{\rho}
+n^{\nu}\nabla_{\mu} \; .
\end{equation}
By virtue of the Dirichlet boundary conditions (1.16), ghost modes
for gravitons are thus found to vanish {\it everywhere}. Hence the
whole ghost sector of simple supergravity yields a vanishing
contribution to the one-loop divergence, when the axial gauges
(1.18) and (2.3), (2.4) are imposed.

Our investigation is part of the more general quantization program
of field theories in noncovariant gauges [50].
For a thorough treatment of supergravity in the axial gauge in
the absence of boundaries, we refer the reader to Ref. [51],
where the gauge-fixing term and the corresponding Faddeev-Popov
ghost terms are studied in detail using tetrad notation,
$\gamma$-matrices and the auxiliary fields. 

\section{Graviton and gravitino modes}

In the Faddeev-Popov path integral for the semiclassical amplitudes
of simple supergravity, one has to add a gauge-averaging term to
the original Euclidean action for gravitons and gravitinos,
jointly with the corresponding ghost term. In Ref. [22] it has
been shown that, in the axial gauge, gravitational perturbations
corresponding to the eigenvalue $\lambda$ obey the equations
\begin{eqnarray}
\lambda \nabla^{\nu}h_{(\lambda)\mu \nu}&=&{1\over 2\alpha}
\Bigr[(K_{\; \; \mu}^{\rho} \; n^{\sigma}
+K^{\rho \sigma} \; n_{\mu})h_{(\lambda)\rho \sigma}
+n_{\mu}n^{\sigma}\nabla^{\rho}h_{(\lambda)\rho \sigma}\Bigr] 
\nonumber\\
&+&{1\over 2\alpha}\Bigr[({\mbox {Tr}}K)
n^{\sigma} \; h_{(\lambda)\mu \sigma}
+n^{\sigma}n^{\rho}\nabla_{\rho}h_{(\lambda)\mu \sigma}\Bigr] \; ,
\end{eqnarray}
\begin{equation}
\left(-\Box +{\lambda \over 2}\right)
h_{(\lambda)\rho}^{\; \; \; \; \rho}
={1\over 2\alpha}n^{\mu}n^{\nu}h_{(\lambda)\mu \nu} \; ,
\end{equation}
subject to the boundary conditions according to which the whole
set of metric perturbations vanishes at the boundary. A 
{\it unique}, smooth and analytic 
solution of such an elliptic boundary-value problem exists,
which picks out transverse-traceless perturbations [22]. The
three-sphere boundary plays indeed an important role, since the
unperturbed extrinsic-curvature tensor $K$ in Eq. (3.1) turns out
to be proportional to the induced three-metric in such a case.
Thus, the $\zeta(0)$ contribution from gravitons is the one first
obtained in Ref. [52]:
\begin{equation}
\zeta_{TT}(0)=-{278\over 45} \; .
\end{equation}

Moreover, in flat Euclidean four-space, the square of the Dirac
operator on gravitinos is $-\Box \; \delta_{\; \; \nu}^{\mu}$.
Thus, denoting by $\psi_{(\lambda)}^{\mu}$ the gravitino
eigenfunctions belonging to the eigenvalue $\lambda$, by 
$n^{\mu}$ the normal to the boundary, and bearing in mind that
the axial gauge-averaging term for gravitinos can be written
as ${1\over 2\alpha}{\widetilde \psi}_{\nu} \; n^{\nu}n^{\sigma}
\; \psi_{\sigma}$ [51], one finds an eigenvalue equation for
gravitino perturbations in the form
\begin{equation}
\left(-\Box \; \delta_{\; \; \nu}^{\mu}
+{1\over \alpha}n^{\mu}n_{\nu}\right)
\psi_{(\lambda)}^{\nu}=\lambda \; \psi_{(\lambda)}^{\mu} \; .
\end{equation}
Covariant differentiation of (3.4), and its contraction with
flat-space $\gamma$-matrices, yield respectively
\begin{equation} 
\nabla_{\mu} \left(-\Box \psi_{(\lambda)}^{\mu}\right)
+{1\over \alpha}\left[({\mbox {Tr}} \; K)
n_{\sigma}\psi_{(\lambda)}^{\sigma}
+n^{\mu}K_{\mu \sigma} \psi_{(\lambda)}^{\sigma}
+n^{\mu}n_{\sigma}\nabla_{\mu}\psi_{(\lambda)}^{\sigma}\right]
=\lambda \nabla_{\mu} \psi_{(\lambda)}^{\mu} \; ,
\end{equation}
\begin{equation}
-\Box \left(\gamma_{\mu}\psi_{(\lambda)}^{\mu}\right)
+{1\over \alpha}\gamma_{\mu}n^{\mu} \left(n_{\sigma}
\psi_{(\lambda)}^{\sigma}\right)=\lambda \; 
\gamma_{\mu}\psi_{(\lambda)}^{\mu} \; .
\end{equation}
Of course, since the normal to the boundary can always be
chosen in the form $n^{\mu}=(1,0,0,0)$, the term 
$n^{\mu}K_{\mu \sigma}$ vanishes, because $K_{00}=K_{0i}=0$.
Thus, denoting by $\Lambda$ the (positive) eigenvalues of the
operator $-\Box$ [22], and by $T_{\lambda}$ the trace
$\gamma_{\mu}\psi_{(\lambda)}^{\mu}$, the system (3.5), (3.6) 
reduces in flat space to
\begin{equation}
-\Box \nabla_{\mu} \psi_{(\lambda)}^{\mu}
+{1\over \alpha}\left[({\mbox {Tr}} \; K)
n_{\sigma}\psi_{(\lambda)}^{\sigma}+n^{\mu}n_{\sigma}
\nabla_{\mu}\psi_{(\lambda)}^{\sigma}\right]
=\lambda \; \nabla_{\mu}\psi_{(\lambda)}^{\mu} \; ,
\end{equation}
\begin{equation}
(\Lambda-\lambda)T_{\lambda}+{1\over \alpha}\gamma_{\mu}
n^{\mu} n_{\sigma}\psi_{(\lambda)}^{\sigma}=0 \; .
\end{equation}
In our flat background, the term on the left-hand side of (3.7)
involving covariant derivatives of $\psi_{(\lambda)}^{\sigma}$
reduces to ${\partial \over \partial \tau}\psi_{(\lambda)}^{0}$.
If we now express $n_{\sigma}\psi_{(\lambda)}^{\sigma}$ in
terms of $T_{\lambda}$ and $\gamma_{\mu}n^{\mu}$ from (3.8),
and insert it into the left-hand side of (3.7), we find that
this reduces to a linear functional of 
$\nabla_{\mu}\psi_{(\lambda)}^{\mu}$ and $T_{\lambda}$.
Thus, a solution of (3.7), (3.8) exists which picks out 
transverse-traceless perturbations which obey the gauge condition
everywhere, i.e., $\forall \tau \in [0,a]$. Moreover, uniqueness
of the elliptic boundary-value problem with spectral boundary 
conditions implies that this is the only possible solution.
More precisely, since we are studying a covariant quantization
scheme, in that the operator in Eq. (3.4) is a four-dimensional
elliptic operator on both normal and tangential components
of Rarita-Schwinger potentials, the full set of spectral boundary
conditions consists of (1.10), (1.11), jointly with nonlocal
boundary conditions of the spectral type on $\psi_{0}^{A}$ and
${\widetilde \psi}_{0}^{A'}$, i.e., 
\begin{equation}
\Bigr[\psi_{0}^{A}(+)\Bigr]_{\partial M}=0 \; ,
\end{equation}
\begin{equation}
\Bigr[{\widetilde \psi}_{0}^{A'}(+)\Bigr]_{\partial M}=0 \; .
\end{equation}
Equations (3.9) and (3.10) involve half of the spin-${1\over 2}$
modes, i.e., those multiplying harmonics on the three-sphere
having positive eigenvalues of the intrinsic three-dimensional
Dirac operator [26,31,45,47]. It should be stressed that boundary
conditions on normal components of spin-${3\over 2}$ potentials
are necessary in any covariant quantization scheme (see also Sec.
IV). By contrast, the analysis in Ref. [26], where reduction to
the physical degrees of freedom was performed before quantization,
did not need (3.9) and (3.10), but used gauge conditions and
linearized supersymmetry constraints.

The resulting gravitino modes contribute [26,31]
\begin{equation}
\zeta_{{3\over 2}}(0)=-{289\over 360} \; .
\end{equation}
By virtue of (2.32), (3.3) and (3.11), the full $\zeta(0)$ value
for simple supergravity in the axial gauge 
is (bearing in mind the anticommuting 
nature of gravitinos)
\begin{equation}
\zeta(0)=-{278\over 45}+{289\over 360}=-{43\over 8} \; ,
\end{equation}
which coincides with the value first obtained in Ref. [53], 
where a simplified analysis was presented, without appreciating
the vanishing contribution of both sets of ghost modes.

No extra contributions occur in Eq. (3.12), since the field
equations for auxiliary fields are algebraic rather than
differential, and as our calculations concern only
one-loop effects, we can impose such field equations at this
point and ignore higher-order ghost contributions [54].
Note also that, instead of (2.3) and (2.4), we might have 
chosen our gauge condition in the form
\begin{equation}
\Gamma^{(AC)C'} \; {_{e}n_{CC'}}=0 \; ,
\end{equation}
\begin{equation}
\gamma^{(A'C')C} \; {_{e}n_{CC'}}=0 \; .
\end{equation}
The resulting ghost operators are a bit more involved,
and the problem deserves careful consideration to check
coincidence of $\zeta(0)$ values.

\section{Luckock-Moss-Poletti boundary conditions}

In Ref. [20] it was proved in detail that, on imposing the LMP
boundary conditions (1.14) and (1.20)--(1.23), the full
$\zeta(0)$ value for gravitons, including gauge and ghost
modes, is 
\begin{equation}
\zeta(0)_{\mbox {gravitons}}=-{758\over 45} \; .
\end{equation}
This result was obtained by using the analytic algorithm of
Ref. [13], and it was confirmed by using the recently corrected
geometric formulas for the asymptotic heat kernel with mixed
boundary conditions [24,55]. Note that the result (4.1) differs
from (3.1), and hence there is no cancellation 
of contributions of gauge and ghost
modes for gravitons with LMP boundary conditions, on using the 
de Donder gauge-averaging functional
$\Phi_{\mu}^{dD} \equiv \nabla^{\nu}\Bigr(h_{\mu \nu}
-{1\over 2}g_{\mu \nu}g^{\rho \sigma}h_{\rho \sigma}\Bigr)$ [20].

We now study the LMP boundary conditions (1.19) for gravitino
perturbations. For this purpose, it is useful to describe some 
basic properties of the geometric theory of heat-kernel
asymptotics, following Refs. [20,24,55].

The Schwinger-DeWitt coefficient $A_{2}$
for the elliptic operator
$-D_{\mu}D^{\mu} + X$, 
where $D_{\mu}$ is a gauge derivative with curvature $F_{\mu \nu}$,
can be written as 
\begin{equation}
16 \pi^{2} A_{2} = \int_{M} b_{2}\;  dV 
+\int_{\partial M}
c_{2}\;  d\Sigma \; .
\end{equation}
The volume coefficient $b_{2}$ is well known [56], while surface terms
depend upon the choice of boundary conditions. We use mixtures
of Dirichlet and Robin boundary conditions,
\begin{mathletters}
\begin{equation}
\Bigr[P_{-}\phi \Bigr]_{\partial M} = 0 \; ,
\end{equation}
\begin{equation}
\Bigr[(\psi 
+ n^{\sigma} \nabla_{\sigma}) P_{+}\phi \Bigr]_{\partial M} 
= 0 \; ,
\end{equation}
\end{mathletters}
where $P_{\pm}$ are projection operators [22,24]. 
The results can be expressed in terms of polynomials in the curvature
tensor $R_{\mu\nu\alpha\beta}$ of the background four-manifold and
in terms of the extrinsic-curvature tensor of the boundary 
(hereafter $R$ is the trace of the Ricci tensor), i.e.,  
\begin{eqnarray}
q& \equiv &  
{8 \over 3} ({\mbox {Tr}}K)^{3} 
+ {16 \over 3} {\mbox {Tr}}K^{3}
-8({\mbox {Tr}}K)({\mbox {Tr}}K^{2})+4R({\mbox {Tr}}K) 
\nonumber\\
&-& 8 R_{\mu\nu}
(({\mbox {Tr}}K) n^{\mu} n^{\nu} + K^{\mu\nu}) + 8 R_{\mu\nu\alpha\beta}
K^{\mu\alpha} n^{\nu} n^{\beta} \; ,
\end{eqnarray}
and
\begin{equation}
p \equiv {\mbox {Tr}}K^{3}-({\mbox {Tr}}K)({\mbox {Tr}}K^{2})
+{2\over 9}({\mbox {Tr}}K)^{3} \; .
\end{equation}
For Dirichlet boundary conditions [24],
\begin{equation}
c_{2}^{D} = {\mbox {Tr}} \biggr[-{1\over 360} q + {2\over 35} p -
{1\over 3} \Bigr(X - {1\over 6} R \Bigr) ({\mbox {Tr}}K) 
- {1\over 2} n^{\sigma} \nabla_{\sigma} 
\Bigr(X- {1 \over 6} R \Bigr) 
+ {1\over 15} C_{\mu\nu\alpha\beta} K^{\mu\alpha}
n^{\nu} n^{\beta} \biggr] \; ,
\end{equation}
while, for Robin boundary conditions [24],
\begin{eqnarray}
c_{2}^{R}&=& {\mbox {Tr}} \biggr[-{1\over 360} q + {2\over 45} p -
{1\over 3} \Bigr(X - {1\over 6} R \Bigr) ({\mbox {Tr}}K) 
+ {1 \over 2} n^{\sigma} \nabla_{\sigma} 
\Bigr(X- {1\over 6} R \Bigr) 
- {4\over 3} \Bigr(\psi - {1\over 3}({\mbox {Tr}}K) \Bigr)^{3}
\nonumber\\
&+& 2 \Bigr(X - {1\over 6}R \Bigr) \psi 
+ \Bigr(\psi - {1\over 3}({\mbox {Tr}}K) \Bigr)
\Bigr({2\over 45} ({\mbox {Tr}}K)^{2} 
- {2\over 15} ({\mbox {Tr}}K^{2}) \Bigr) 
\nonumber\\
&+& {1\over 15} C_{\mu\nu\alpha\beta} K^{\mu\alpha}
n^{\nu} n^{\beta} \biggr] \; .
\end{eqnarray}
For mixed boundary conditions [24,55],
\begin{eqnarray}
c_{2}&=& {\mbox {Tr}} \Bigr[P_{+} c_{2}^{R} 
+ P_{-} c_{2}^{D} -{2\over 15}
(P_{+;i})(P_{+}^{;i}) ({\mbox {Tr}}K) 
- {4\over 15} (P_{+;i})(P_{+;j})K^{ij}
\nonumber\\
&+& {4\over 3} (P_{+;i})(P_{+}^{;i}) P_{+} \psi 
- {2\over 3} P_{+}
(P_{+}^{;i}) n^{\mu} F_{i\mu}\Bigr] \; ,
\end{eqnarray}
where Latin indices run from 1 to 3 and Greek indices run
from 0 to 3 [20,24,55].

Following Ref. [57] we now consider the gravitino action  
in the Euclidean form
\begin{equation}
S_{3/2}  = {1 \over 2} \int_{M} d^{4}x \sqrt{g}
\; \varepsilon^{\alpha\beta\gamma\delta}
\tilde{\psi}_{\alpha}\gamma_{5}\gamma_{\beta}\nabla_{\gamma}
\psi_{\delta} \; ,
\end{equation}
where torsion terms are neglected since we are interested in
one-loop calculations [31,37,54].
To fix the gauge condition we use the change of variables
\begin{equation}
\psi_{\mu} \equiv \kappa_{\mu}-{1 \over 2}\gamma_{\mu}\gamma^{\alpha}
\kappa_{\alpha} \; .
\end{equation}
Hence the gauge-fixed action including ghosts becomes [54]
\begin{equation}
S_{3/2} = \int_{M} d^{4}x \sqrt{g} \biggr[ 
{1\over 2}\tilde{\kappa}_{\mu}{\widehat \nabla}\kappa^{\mu} +
\tilde{\epsilon}{\widehat \nabla}\epsilon
+{1\over 2}\tilde{\eta}{\widehat \nabla} \eta \biggr] \; ,
\end{equation}
where the {\it hat} denotes contraction with $\gamma$-matrices
according to the standard rule
${\widehat \nabla} \equiv \gamma_{\rho}\nabla^{\rho}$.
Here, $\epsilon,{\tilde \epsilon}$ are the usual 
Faddeev-Popov ghost fields while $\eta$
is the Nielsen-Kallosh ghost field [58,59] (see also Appendix A).
All these ghost fields behave as bosonic fields, since gravitinos
have a Grassmann nature [35,54,58,59]. To evaluate
the Schwinger-DeWitt coefficient $A_{2}$ for fermionic
fields we use the familiar identity
$\log \det A = {1 \over 2} \log \det (A^{+} A)$
for the Dirac operator $A$. This makes it possible to use
squared fermionic operators, which reduce to the standard 
Laplace operator in flat Euclidean four-space.

Let us now define the projectors for spin-1/2 and spin-3/2 fields.
Projectors for a spin-1/2 field have the following form:
\begin{equation}
P_{\pm} \equiv {1\over 2}(1 \pm i \gamma_{5} \hat{n}) \; ,
\end{equation}
while the (mixed) boundary conditions for this field are [57] 
\begin{mathletters}
\begin{equation}
\Bigr[P_{-}\phi\Bigr]_{\partial M} = 0 \; , 
\end{equation}
\begin{equation}
\left[\left({1 \over 2} {\mbox {Tr}} K 
+ n_{\alpha} \nabla^{\alpha} \right)P_{+}\phi\right]_{\partial M} 
= 0 \; .
\end{equation}
\end{mathletters}
Note that the square of our $\gamma_{5}$ is 1, to ensure that
(4.12) is a projector. Thus, since the square of $(\gamma_{0})_{E}$,
the Euclidean $\gamma_{0}$, is also 1, one has to take 
$$
\gamma_{5}=(\gamma_{0})_{E}\gamma_{1}\gamma_{2}\gamma_{3}
=i(\gamma_{0})_{L}\gamma_{1}\gamma_{2}\gamma_{3}
=(\gamma_{5})_{L} \; ,
$$
i.e., $\gamma_{5}$ coincides with its Lorentzian counterpart.

For a spin-3/2 field we have the following projectors [57]:
\begin{equation}
\Bigr({\cal P}_{\pm}\Bigr)_{\; \nu}^{\mu} 
\equiv Q^{\mu}_{\nu}P_{\mp} + n^{\mu}n_{\nu}
P_{\pm} \; ,
\end{equation}
where $Q^{\mu}_{\nu} \equiv 
\delta^{\mu}_{\nu} - n^{\mu}n_{\nu}$ [22], 
and $P_{\pm}$ are the projectors defined in (4.12).
The LMP boundary conditions for the massless spin-3/2
field are [57]
\begin{equation}
\Bigr[{\cal P}_{-}\kappa_{\mu}\Bigr]_{\partial M} = 0 \; ,
\end{equation}
\begin{equation}
\left[\Bigr({1 \over 2}({\mbox {Tr}}K)\delta^{\mu}_{\nu} + i \gamma_{5}
{\widehat K}^{\mu} n_{\nu} 
+ i \gamma_{5} n^{\mu} {\widehat K}_{\nu} +
n_{\alpha}\nabla^{\alpha}\delta^{\mu}_{\nu}\Bigr) {\cal P}_{+}
\kappa_{\mu}\right]_{\partial M} = 0 \; ,
\end{equation}
where ${\widehat K}^{\mu} \equiv K^{\mu\nu}\gamma_{\nu}$.
The LMP choices (4.13), (4.15), (4.16), jointly with (1.14),
(1.20)--(1.23), correspond to boundary conditions which are
invariant under the BRST form of the supersymmetry transformations,
{\it with the exception} of 
the spatial components $h_{ij}$ and $\psi_{i}^{A}$ [57].
Thus, one should bear in mind that the LMP boundary conditions
are not completely gauge-invariant [20,22]. Note also that the
spatial component of (4.15) corresponds to (1.19), while (4.16)
can be obtained by requiring that (4.15) should be preserved
under the action of the Dirac operator on $\kappa_{\mu}$ [31,57]. 

Now one can easily find that in our flat four-manifold with a
three-sphere boundary the value of $q$ in Eq. (4.4) is
(hereafter, the three-sphere radius is set to 1 for simplicity,
since its effect is cancelled by the integration over the boundary
in (4.2))
\begin{equation}
q = 16 \; ,
\end{equation}
while Eq. (4.5) leads to
\begin{equation}
p = 0 \; .
\end{equation}
Insertion of Eqs. (4.17) and (4.18) into Eq. (4.6) leads to
\begin{equation}
c_{2}^{D} = - {2 \over 45} ({\mbox {Tr}}I) \; .
\end{equation}
With our notation, $({\mbox {Tr}}I)$ involves traces over both
spinor and tensor indices, and is equal to 8 for 
(massless) gravitinos. Moreover,
from Eq. (4.7) one finds for the case of Robin boundary conditions
\begin{equation}
c_{2}^{R} ={58\over 45} ({\mbox {Tr}}I) 
- {4\over 3} ({\mbox {Tr}} \Omega^{3})
+ 4 ({\mbox {Tr}} \Omega^{2}) - 4 ({\mbox {Tr}}\Omega) \; ,
\end{equation}
where we have defined 
$\Omega_{\; \nu}^{\mu} \equiv {1\over 2}({\mbox {Tr}}K)
\delta_{\; \nu}^{\mu}+i\gamma_{5}{\widehat K}^{\mu}n_{\nu}
+i\gamma_{5}n^{\mu}{\widehat K}_{\nu}$. Thus, in the case of
gravitinos, $\Omega_{\; \nu}^{\mu}$ is not proportional to the
identity, and one has to calculate the following traces:
\begin{equation}
{\mbox {Tr}} \Omega^{\mu}_{\; \mu} 
= 2 ({\mbox {Tr}}K) ({\mbox {Tr}}1) = 6({\mbox {Tr}}1) \; ,
\end{equation}
\begin{equation}
{\mbox {Tr}} \Omega^{\mu}_{\; \nu}\Omega^{\nu}_{\; \mu} =
\Bigr[({\mbox {Tr}} K)^{2} + 2 ({\mbox {Tr}}K^{2})\Bigr] 
({\mbox {Tr}}1) = 15 ({\mbox {Tr}}1) \; ,
\end{equation}
\begin{equation}
{\mbox {Tr}} \Omega^{\mu}_{\; \nu}\Omega^{\nu}_{\; \rho}
\Omega^{\rho}_{\; \mu}
=\left[{1\over 2} ({\mbox {Tr}} K)^{3} 
+ 3 ({\mbox {Tr}}K) ({\mbox {Tr}}K^{2})
\right] ({\mbox {Tr}}1) 
= {81 \over 2} ({\mbox {Tr}}1) \; ,
\end{equation}
where $({\mbox {Tr}}1)$ denotes instead the trace over spinor
indices only, and is equal to $2={1\over 4}({\mbox {Tr}}I)$ in
our case. Substituting Eqs. (4.21)--(4.23) 
into Eq. (4.20) one has
\begin{equation}
c_{2}^{R} = - {289\over 90} ({\mbox {Tr}}I) \; .
\end{equation}
To evaluate $c_{2}$, we also need the spatial components 
of the four-dimensional covariant
derivative of the projector ${\cal P}_{+}$:
\begin{equation}
{\Bigr({\cal P}_{+}\Bigr)_{\; \nu}^{\mu}}_{;j} 
= i\gamma_{5}{\hat k}_{j} \left(n^{\mu}n_{\nu}-
{1\over 2}\delta^{\mu}_{\nu}\right) +
i\gamma_{5}{\hat n}\Bigr(K^{\mu}_{j}n_{\nu}
+ n^{\mu} K_{\nu j}\Bigr) \; .
\end{equation}
After substitution of Eqs. (4.19), (4.24) and (4.25) into
Eq. (4.8) one obtains
\begin{equation}
c_{2} = -{169\over 90} ({\mbox {Tr}}I) \; .
\end{equation}
Bearing in mind that the gravitino has 8 degrees of freedom one
thus finds from (4.2) and (4.26)
\begin{equation}
{A_{2}}_{\mbox {gravitino}} = -{169\over 90} \; .
\end{equation}
We know also that the value of $\zeta(0)$ for a massless 
spin-${1\over 2}$ field subject to the local boundary 
conditions (4.13) is [24--29]
\begin{equation}
\zeta_{1/2}(0) ={11 \over 360} \; .
\end{equation}
Thus, combining (4.27), (4.28), and bearing in mind that
gravitinos are fermionic while their Faddeev-Popov and
Nielsen-Kallosh ghosts are bosonic, one finds
\begin{equation}
\zeta_{3/2}(0) 
= -{169\over 90}-{11\over 120}=-{709\over 360} \; ,
\end{equation}
which leads to
\begin{equation}
\zeta(0)_{N=1 \; {\mbox {SUGRA}}}
=-{758\over 45}+{709\over 360}=-{119\over 8} \; ,
\end{equation}
by virtue of (4.1). Thus, simple supergravity is not even
one-loop finite in the background with boundary motivated
by quantum cosmology [16--31]. Moreover, one can check that,
with a naive combination of contributions of different fields
in extended supergravity theories, one again fails to obtain
exact cancellations of one-loop divergences (cf. Ref. [57]).
However, our own understanding of antisymmetric tensor fields
and curved backgrounds with boundary for higher-N supergravity
models remains insufficient, and hence this appears as an
open problem for further research (see also Sec. VI).

\section{Two-boundary problem}

So far we have only studied backgrounds motivated by quantum
cosmology, i.e., where one three-surface shrinks to zero,
so that only one bounding three-sphere occurs [4,5]. In quantum
field theory, however, one deals with a path-integral representation
of the propagation amplitude with suitable data on {\it two}
boundary three-surfaces [15]. Thus, this section is devoted to
the one-loop analysis of simple supergravity when two concentric
three-sphere boundaries occur, with radii $\tau_{-}$ and
$\tau_{+}$ respectively ($\tau_{+}> \tau_{-}$). As in Secs. II
and III, we use the axial gauge with spectral boundary conditions
for gravitinos, and completely gauge-invariant
boundary conditions for metric perturbations [20,22,32].

We know from Ref. [22] and Sec. III that, in the axial gauge,
the only surviving graviton modes are transverse-traceless.
Hence they contribute [19]
\begin{equation}
\zeta(0)_{\mbox {gravitons}}=\zeta_{TT}(0)=-5 \; ,
\end{equation}
while ghost modes for gravitons vanish everywhere [22]. 
Moreover, the gravitino modes picked out by the axial gauge
take the form [26,31]
\begin{mathletters}
\begin{equation}
f_{n}(\tau)= \sqrt{\tau}I_{n+2}(M\tau)
+ \beta_{n} \sqrt{\tau}K_{n+2}(M\tau) \; ,
\end{equation}
\begin{equation}
{\widetilde f}_{n}(\tau)= \sqrt{\tau}I_{n+3}(M\tau)
- \beta_{n} \sqrt{\tau}K_{n+3}(M\tau) \; ,
\end{equation}
\end{mathletters}
where $\beta_{n}$ are 
some constants, and the integer $n$ is
$\geq 0$. The modes $f_{n}$ should vanish at $\tau=\tau_{+}$,
and the modes ${\widetilde f}_{n}$ should vanish at
$\tau=\tau_{-}$. 
The full degeneracy with spectral boundary conditions
is then $2(n+1)(n+4)$, and the equation obyed by the eigenvalues
by virtue of these boundary conditions is
\begin{equation}
I_{n+2}(M\tau_{+})K_{n+3}(M\tau_{-})+I_{n+3}(M\tau_{-})
K_{n+2}(M\tau_{+})=0 \; .
\end{equation}
We can now apply the technique described in Ref. [13] and used
several times by the authors in recent work 
[17--21,23,27--29]. Thus, many details can be omitted, and we
limit ourselves to say that $\zeta(0)$ for gravitinos has the
general structure [13]
\begin{equation}
\zeta(0)=I_{\mbox {log}}+I_{\mbox {pole}}(\infty)
-I_{\mbox {pole}}(0) \; ,
\end{equation}
where
\begin{equation}
I_{\mbox {log}}= 0 \; ,
\end{equation}
\begin{equation}
I_{\mbox {pole}}(\infty)=0 \; ,
\end{equation}
\begin{equation}
I_{\mbox {pole}}(0)= 0 \; ,
\end{equation}
which implies
\begin{equation}
\zeta(0)_{\mbox {gravitinos}}= 0 \; .
\end{equation}

The ghost contribution to $\zeta(0)$ for gravitinos is obtained
by setting to zero at the boundaries the first derivatives of
the modes (2.21)--(2.24). Bearing in mind that the roles of
twiddled and untwiddled modes are interchanged with respect to 
the spin-${1\over 2}$ analysis of Ref. [47], spectral boundary
conditions for ghost modes can be written as follows:
\begin{equation}
{dm_{np}\over d\tau}(\tau=\tau_{-})=0 \; ,
\end{equation}
\begin{equation}
{dr_{np}\over d\tau}(\tau=\tau_{-})=0 \; ,
\end{equation}
\begin{equation}
{d{\widetilde m}_{np}\over d\tau}(\tau=\tau_{+})=0 \; ,
\end{equation}
\begin{equation}
{d{\widetilde r}_{np}\over d\tau}(\tau=\tau_{+})=0 \; .
\end{equation}
Equations (5.9) and (5.12) lead to incompatible solutions for
$\lambda_{n}$, i.e., $\lambda_{n}=-{n\over \tau_{-}} \leq 0$
and $\lambda_{n}={(n+3)\over \tau_{+}} > 0$, while Eqs. (5.10)
and (5.11) lead to incompatible solutions for ${\tilde \lambda}_{n}$,
i.e., ${\tilde \lambda}_{n}=-{(n+3)\over \tau_{-}}<0$ and
${\tilde \lambda}_{n}={n\over \tau_{+}} \geq 0$. Thus, the boundary
conditions (5.9)--(5.12) imply that no non-trivial ghost modes exist,
and by virtue of Eqs. (5.1) and (5.8) the full $\zeta(0)$ value
in the axial gauge in the two-boundary problem is found to be 
\begin{equation}
\zeta(0)_{N=1 \; {\mbox {SUGRA}}}=-5 \; . 
\end{equation}

\section{Concluding remarks}

This paper has evaluated one-loop divergences in simple supergravity
for two choices of mixed boundary conditions. First, on
imposing Barvinsky boundary conditions for gravitons in the
axial gauge (1.18), jointly with spectral boundary conditions
for gravitinos in the axial gauge (2.3) and (2.4), the full
one-loop divergence has been found in Eq. (3.12). This has been
achieved by means of a two-spinor analysis of the ghost operators
(2.5) and (2.6), jointly with a nontrivial cancellation resulting
from the $\zeta$-functions (2.28) and (2.29). This calculation
provides a very good example of the role played by finite-dimensional
null-spaces in one-loop calculations [48]. The analysis of the 
ghost sector for gravitons in the axial gauge [22] has also 
been used.

Second, Luckock-Moss-Poletti (LMP) boundary conditions (1.14),
(1.20)--(1.23), (4.13), (4.15) and (4.16)
have been studied by means of the geometric
formulas for the asymptotic heat kernel with mixed boundary
conditions, in the corrected form obtained in Refs. [24,55].
Unlike the boundary conditions studied in Secs. II and III,
LMP boundary conditions do not lead to a vanishing contribution
of ghost and gauge modes, and the full one-loop divergence
is the one given in Eq. (4.30).

Third, the two-boundary problem in the axial gauge for 
supergravity with spectral boundary conditions has been
studied. The resulting one-loop divergence has been obtained
in Eq. (5.13).

For pure gravity, the investigation of one-loop divergences
in the absence of boundaries was already a formidable task.
It was indeed a great achievement of Ref. [60] to calculate
all one-loop divergences of pure gravity and of gravitation
interacting with scalar particles. In particular, in the former
case, it was found in Ref. [60] that no physically relevant
divergences remain, since they can all be absorbed in a field
renormalization. In quantum supergravity, one might hope that
the supersymmetry relating bosonic and fermionic fields
improves the finiteness properties [1,31,35,61], and the 
perturbative analysis becomes more involved. One has then to
consider S-matrix elements, topological invariants and
boundary effects [1,30,31]. As far as we know, our paper has
presented the first detailed investigation of one-loop effects
in the axial gauge with spectral boundary conditions. More
generally, the results (3.12), (4.30) 
and (5.13) seem to point out that
simple supergravity is not even one-loop finite in 
backgrounds with boundary. 

Note that, despite the singularity at the origin of our flat
background with boundary
in Secs. II--IV, the axial gauge leads to well
defined calculations since the resulting operators on
perturbative modes reduce to the standard covariant operators
(e.g. the Laplacian on transverse-traceless tensors). 
Interestingly, since the axial
gauge-averaging functional selects transverse-traceless
perturbations, whereas all remaining gauge modes are forced
to vanish everywhere, no exact cancellation of all contributions
to the one-loop divergence seems to occur in this two-boundary
problem. This is quite different from the result found in
non-covariant gauges for Euclidean Maxwell theory [17], where
the full $\zeta(0)$ value was found to vanish. Thus, more work
might be in order to interpret correctly the value obtained in
Eq. (5.13). 

Moreover, we do not yet know how to evaluate explicitly higher-order 
effects in perturbation theory in the presence of boundaries
(cf. Refs. [61,62]), neither in quantum cosmological backgrounds
(where one boundary three-surface shrinks to zero [4,5]),
nor in the familiar field-theoretical case where both boundary
three-geometries are actually present. Although higher-order
calculations cannot be performed by means of $\zeta$-function
techniques, the heat kernel provides, in principle, a 
nondiagrammatic way to study higher orders. Another relevant
problem is given by one-loop calculations in the presence of
boundaries for higher-N supergravity models. These are naturally
formulated on backgrounds with a nonvanishing cosmological 
constant [63--65], and the analytic or geometric features of the 
one-loop calculation remain unknown.

The form of the graviton and gravitino propagators in the axial
gauge with Barvinsky [32] and spectral [26,31] boundary conditions
respectively is also unknown, and deserves careful consideration,
jointly with the analysis of the full one-loop effective action
[66--68]. 

A further interesting issue might be the re-formulation of our
perturbative analysis within the framework proposed in Ref. [69],
where an effective action for gauge fields and gravitation has
been constructed that is gauge-invariant and independent of the
choice of gauge-averaging term, in such a way that new Feynman
rules without ghost lines are obtained for the first time.

Thus, many exciting open problems remain, and this seems to add
evidence in favor of no end being in sight for cosmology and
fundamental physics [31].

\acknowledgements
One of us (G.E.) would like to thank Peter D'Eath for suggesting
the problem and constant encouragement over nine years. 
Scientific collaboration with Ivan Avramidi on the axial gauge
in Euclidean quantum gravity has been of invaluable help,
and correspondence with Klaus Kirsten was also very helpful.
The work of A.Y.K. was partially supported by the
Russian Foundation for Fundamental Researches through Grant No.
96-02-16220-a, and by the Russian Research Project
``Cosmomicrophysics".
\appendix
\section{}
The massless spinor fields occurring in Eqs. (1.2) and (1.3)
obey the Rarita-Schwinger equations [12,38,70,71]
\begin{equation}
\epsilon^{B'C'} \; 
\nabla_{A(A'} \; \gamma_{\; \; B')C'}^{A}=0 \; ,
\end{equation}
\begin{equation}
\nabla^{B'(B} \; \gamma_{\; \; \; B'C'}^{A)} =0 \; ,
\end{equation}
\begin{equation}
\epsilon^{BC} \; 
\nabla_{A'(A} \; \Gamma_{\; \; \; B)C}^{A'}=0 \; ,
\end{equation}
\begin{equation}
\nabla^{B(B'} \; \Gamma_{\; \; \; \; BC}^{A')}=0 \; .
\end{equation}
Thus, by virtue of Eq. (1.6), the second potential $\rho$ obeys
the equation [12]
\begin{equation}
\epsilon_{FL}\nabla_{AA'}\nabla^{B'(F} \; 
\rho_{B'}^{\; \; \; A)L}
+{1\over 2}\nabla_{\; \; A'}^{A} \nabla^{B'M}
\rho_{B'(AM)}+{\Box}_{AM} \; \rho_{A'}^{\; \; \; (AM)}
+{3\over 8}\Box\rho_{A'}=0 \; ,
\end{equation}
where ${\Box}_{AM} \equiv \nabla_{B'(A} \;
\nabla_{M)}^{\; \; \; B'}$, and $\Box \equiv
\nabla^{CF'}\nabla_{CF'}$.
Remarkably, if the following first-order equation holds 
[12,38,39,72]: 
\begin{equation}
\nabla^{B'(F} \; \rho_{B'}^{\; \; \; A)L}=0 \; ,
\end{equation}
then (A5) becomes an identity. Thus, the analysis of second
potentials is reduced to study Eq. (A6) jointly with [12,39]
\begin{equation}
\nabla^{B(F'} \; \theta_{B}^{\; \; A')L'}=0 \; .
\end{equation}
Equations (A6) and (A7) are invariant under the infinitesimal 
gauge transformations
\begin{equation}
{\widehat \rho}_{B'}^{\; \; \; AL} \equiv \rho_{B'}^{\; \; \; AL}
+\nabla_{B'}^{\; \; \; A} \; \omega^{L} \; ,
\end{equation}
\begin{equation}
{\widehat \theta}_{B}^{\; \; A'L'} \equiv \theta_{B}^{\; \; A'L'}
+ \nabla_{B}^{\; \; A'} \; \sigma^{L'} \; ,
\end{equation}
if and only if Eqs. (1.8) and (1.9) hold. Thus, to have freely
specifiable gauge fields $\omega^{L}$ and $\sigma^{L'}$,
the Ricci-flat background is further restricted to be totally
flat [12].

In our paper, we rely in part on 
the spinor notation described in detail in
Refs. [36,37,47]. Thus, we only briefly describe some basic 
identities used in performing the ghost analysis in Sec. II. 
Denoting by $n^{AA'}$ the two-spinor version of the Lorentzian
normal to the boundary, the corresponding Euclidean normal is
defined as
\begin{equation}
{_{e}n^{AA'}} \equiv -i \; n^{AA'} \; .
\end{equation}
The spinor version of the tetrad $e_{\; \; \mu}^{a}$ is obtained
out of the Infeld-van der Waerden symbols as 
$e_{\; \; \; \; \; \mu}^{AA'} \equiv e_{\; \; \mu}^{a} \;
\sigma_{a}^{\; \; AA'}$ [37]. In terms of these objects, the following
identities are found to hold (cf. Refs. [31,36,37])
\begin{equation}
{_{e}n_{AA'}} \; {_{e}n^{AB'}}=-{1\over 2}
\epsilon_{A'}^{\; \; \; B'} \; ,
\end{equation}
\begin{equation}
{_{e}n^{AA'}} \; e_{AA'}^{\; \; \; \; \; i}=0 \; ,
\end{equation}
\begin{equation}
e_{AA'i} \; e^{BB'i}=-\epsilon_{A}^{\; \; B} \;
\epsilon_{A'}^{\; \; B'}
-{_{e}n_{AA'}} \; {_{e}n^{BB'}} \; ,
\end{equation}
\begin{equation}
{ }^{(4)}\nabla_{i}\nu_{C}={ }^{(3)}\nabla_{i}\nu_{C}
-{1\over \tau} {_{e}n_{CD'}} \; e_{B \; \; \; i}^{\; \; D'}
\; \nu^{B} \; ,
\end{equation}
where ${ }^{(3)}\nabla_{i}$ denotes three-dimensional covariant
differentiation tangentially with respect to the intrinsic 
Levi-Civita connection of the boundary. Our $\nabla^{AA'}$
operator denotes always four-dimensional spinor covariant 
derivative: $\nabla^{AA'} \equiv e_{\; \; \; \; \; \mu}^{AA'}
\; \nabla^{\mu}$, where, in particular, 
$e_{\; \; \; \; \; 0}^{AA'}={_{e}n^{AA'}}$. Note also that,
in our flat background, the Euclidean-time component of
covariant differentiation reduces to partial derivative with
respect to $\tau$, since the shift vectors vanish, and hence
the $\omega_{\; \; \; \; 0}^{AB}$ component of the connection
one-forms vanishes [36].

The unbarred harmonics occurring in Eqs. (2.7) and (2.8) obey,
$\forall n \geq 0$, the eigenvalue equations
\begin{equation}
e^{AA'j} \; { }^{(3)}\nabla_{j}\rho_{A}^{nq}
=-\Bigr(n+{3\over 2}\Bigr){_{e}n^{AA'}} \; 
\rho_{A}^{nq} \; ,
\end{equation}
\begin{equation}
e^{AA'j} \; { }^{(3)}\nabla_{j}\sigma_{A'}^{nq}
=-\Bigr(n+{3\over 2}\Bigr){_{e}n^{AA'}} \; \sigma_{A'}^{nq} \; ,
\end{equation}
\begin{equation}
e^{AA'j} \; { }^{(3)}\nabla_{j}
\Bigr({_{e}n_{\; \; A'}^{B}} \; \rho_{B}^{nq}\Bigr)
=\Bigr(n+{3\over 2}\Bigr) {_{e}n^{AA'}} \;
\Bigr({_{e}n_{\; \; A'}^{B}} \; \rho_{B}^{nq}\Bigr) \; ,
\end{equation}
\begin{equation}
e^{AA'j} \; { }^{(3)}\nabla_{j} 
\Bigr({_{e}n_{A}^{\; \; B'}} \; \sigma_{B'}^{nq}\Bigr)
=\Bigr(n+{3\over 2}\Bigr) {_{e}n^{AA'}}
\Bigr({_{e}n_{A}^{\; \; B'}} \; \sigma_{B'}^{nq}\Bigr) \; .
\end{equation}
The corresponding eigenvalue equations for the barred harmonics
${\overline \rho}_{A'}^{nq}$ and ${\overline \sigma}_{A}^{nq}$ can
be obtained by complex conjugation of the spinors occurring in
(A15)--(A18), when regarded as $SL(2,C)$ spinors. Thus, a sign
change occurs on the right-hand side of the barred counterpart
of (A15)--(A18) by virtue of the $-i$ factor in (A10) and of
reality of the Lorentzian normal, which imply that the Euclidean
normal is anti-Hermitian [47].

The readers who are more familiar with $\gamma$-matrix formalism,
may find it useful to read the following outline of basic
properties. The Rarita-Schwinger potential $\psi_{\mu}$ admits a
unique Hodge-like decomposition [54] into its trace part 
${1\over 4}\gamma_{\mu}\Bigr(\gamma^{\sigma}\psi_{\sigma}\Bigr)$,
a longitudinal part $\psi_{\mu}^{L}$ and a transverse-traceless 
part $\psi_{\mu}^{TT}$, such that
\begin{equation}
\psi_{\mu}=\psi_{\mu}^{TT}+\psi_{\mu}^{L}
+{1\over 4}\gamma_{\mu}\Bigr(\gamma^{\sigma}\psi_{\sigma}\Bigr) \; ,
\end{equation}
where
\begin{equation}
\gamma^{\mu} \; \psi_{\mu}^{L}=0 \; ,
\end{equation}
\begin{equation}
\psi_{\mu}^{\rm L}=\nabla_{\mu}\chi 
-{1\over 4}\gamma_{\mu} \gamma^{\sigma}\nabla_{\sigma}\chi \; ,
\end{equation}
\begin{equation}
\gamma^{\mu} \; \psi_{\mu}^{TT}=0 \; ,
\end{equation}
\begin{equation}
\nabla^{\mu} \psi_{\mu}^{TT}=0 \; ,
\end{equation}
for some spinor field $\chi$ which is determined up to
solutions of the twistor equation [39,54,65]. This
decomposition is orthogonal in that
\begin{equation}
\int_{M}g^{\mu \nu} \Bigr(\psi_{\mu}^{L}\Bigr)^{\dagger}
\psi_{\nu}^{TT} \; \sqrt{g} d^{4}x=0 \; .
\end{equation}

The need for a third ghost in Sec. IV can be understood by
pointing out that a massless Rarita-Schwinger field has 8
degrees of freedom, and only 4 of these are removed by the
Faddeev-Popov ghosts. Hence one is left 
with 4 degrees of freedom, instead
of the 2 which yields the correct counting of helicity states of
a massless particle [58]. More precisely, on imposing the
gauge conditions 
\begin{equation}
\gamma^{\rho}\psi_{\rho}=b \; ,
\end{equation}
\begin{equation}
{\widetilde \psi}_{\lambda}\gamma^{\lambda}={\tilde b} \; ,
\end{equation}
where $\psi_{\rho}$ and ${\widetilde \psi}_{\rho}$ are
independent gauge fields in the Rarita-Schwinger Lagrangian
\begin{equation}
{\cal L}_{cl}^{RS}=-{\widetilde \psi}_{\lambda}
\; \epsilon^{\lambda \mu \nu \rho} \; \gamma_{\mu}\gamma_{5}
D_{\nu}\psi_{\rho} \; ,
\end{equation}
the path-integral representation of the vacuum to vacuum
amplitude turns out to be [58] 
\begin{equation} 
Z={\mbox {const}} \times \int [d\psi][d{\widetilde \psi}]
({\mbox {det}}\; \gamma D)^{-2} \delta \Bigr(\gamma^{\rho}
\psi_{\rho}-b \Bigr) \delta \Bigr({\widetilde \psi}_{\lambda}
\gamma^{\lambda}-{\tilde b}\Bigr) 
{\mbox e}^{-S_{cl}[\psi,{\tilde \psi}]}
\int[dg][d{\overline g}] \; .
\end{equation}
Finally, this is re-expressed as [57,58]
\begin{eqnarray}
Z&=& ({\mbox {const}})' \times \int[d\psi] [d{\widetilde \psi}]
({\mbox {det}} \; \gamma D)^{-2}({\mbox {det}} \; M)^{-1} 
\nonumber\\
&\times & {\mbox {exp}} \biggr \{-S_{cl}[\psi,{\tilde \psi}]
-{1\over 2}\int d^{4}x \sqrt{\mid g \mid}
\; {\widetilde \psi}_{\rho}\gamma^{\rho}\; M \;
\gamma^{\lambda}\psi_{\lambda} \biggr \} \; .
\end{eqnarray}
The nontrivial feature is the occurrence of 
$({\mbox {det}}\; M)^{-1}$, where, in the background gauge, 
$M$ takes the form $\gamma^{\lambda}D_{\lambda}$.

At a deeper level, as shown in detail in Ref. [59], Eq. (A29)
reflects the need for modified Feynman rules in supergravity.
Denoting by ${\cal R}_{\alpha}^{i}$ the generators of local
gauge transformations in supergravity, they turn out to obey
the supercommutation relations [35,59]
\begin{equation}
\left \{ {\cal R}_{\; \alpha,j}^{i} \; , \;
{\cal R}_{\; \beta}^{j} \right \} = {\cal R}_{\; \gamma}^{i}
\; f_{\; \alpha \beta}^{\gamma}
+\eta_{\; \; \; \alpha \beta}^{ij} \; S_{,j} \; ,
\end{equation}
where the $f_{\; \beta \gamma}^{\alpha}$ and
$\eta_{\; \; \; \alpha \beta}^{ij}$ are functions of the field
variables. One then finds that the ghost action contains
quartic ghost couplings (i.e., a term quadratic in the ghost
and quadratic in the anti-ghost). The antighost transforms as
in the standard BRST case, while the gauge fields and the
ghost obey a generalization of the standard BRST transformation [59].
All this leads to modified Ward-Slavnov-Taylor identities which 
ensure the gauge invariance of the quantum theory.

\section{}
In Sec. II, we rely on the Riemann and Hurwitz $\zeta$-functions,
defined respectively as [73]
\begin{equation}
\zeta_{R}(s) \equiv \sum_{n=1}^{\infty}n^{-s} \; ,
\end{equation}
\begin{equation}
\zeta_{H}(s,\alpha) \equiv \sum_{m=0}^{\infty}
(m+\alpha)^{-s} \; .
\end{equation}
These functions have an analytic continuation to the whole
complex-$s$ plane as meromorphic functions.
A very useful formula yields [73]
\begin{equation}
\zeta_{H}(-n,\alpha)=-{B_{n+1}(\alpha)\over (n+1)} \; ,
\end{equation}
where $B_{n}(x)$ denotes the Bernoulli polynomial of order $n$.
In particular, we only need
\begin{equation}
B_{1}(x)=x-{1\over 2} \; ,
\end{equation}
\begin{equation}
B_{2}(x)=x^{2}-x+{1\over 6} \; ,
\end{equation}
\begin{equation}
B_{3}(x)=x^{3}-{3\over 2}x^{2}+{1\over 2}x \; .
\end{equation}
Moreover, we need the following values of the 
$\zeta$-function defined in (B1): $\zeta_{R}(-2)=0,
\zeta_{R}(-1)=-{1\over 12}, \zeta_{R}(0)=-{1\over 2}$ [73].

\end{document}